# MODELTALK: A Framework for Developing Domain Specific Executable Models *

Atzmon Hen-Tov

Pontis Ltd.
Glil Yam 46905, Israel
atzmon@pontis.com

David H. Lorenz

The Open University of Israel
108 Ravutski St., Raanana 43107, Israel
lorenz@openu.ac.il

Lior Schachter

Pontis Ltd.
Glil Yam 46905, Israel
liors@pontis.com

## Abstract

Developing and maintaining complex, large-scale, product line of highly customized software systems is difficult and costly. Part of the difficulty is due to the need to communicate business knowledge between domain experts and application programmers. Domain specific model driven development (MDD) addresses this difficulty by providing domain experts and developers with domain specific abstractions for communicating designs. Most MDD implementations take a generative approach. In contrast, we adopt an interpretive approach to domain specific model driven development. We present a framework, named MODELTALK, that integrates MDD, dependency injection and meta-modeling to form an interpretive, domain specific model modeling framework. The framework is complemented by tool support that provides developers with the same advanced level of usability for modeling as they are accustomed to in programming environments. MODELTALK is used in a commercial setting for developing a product line of Telco grade business support systems (BSS).

*Categories and Subject Descriptors*    D2.6 [*Programming Environments*]: Programmer workbench; D3.2 [*Language Classifications*]: Extensible languages, Object-oriented languages

*General Terms*    Design, Languages

## 1.  Introduction

Modern business application development is complex. It involves several domains of expertise, dealing with both functional and extra-functional requirements, all complicating the communication between domain users and domain experts. Working with domain specific models alleviates some of this complexity by communicating domain abstractions in designs.

In this work, we present a framework, named MODELTALK, for developing domain specific executable models. An *executable model* [17, 12] is a model that drives the execution of the system. The major virtue of an executable model is that changes in the model are automatically reflected in the system [24]. MODELTALK is an interpretive, domain specific modeling framework: the model is the primary source of the system; the desired behavior of the runtime system is achieved by interpreting the model.

MODELTALK integrates the principle of domain driven development [22, 19] with the technique of dependency injection. *Dependency injection* [15, 10] is a mechanism for defining external dependency declaratively (e.g., as an object graph configuration in XML) that can be injected into the runtime system (e.g., into Java objects). The major virtue of dependency injection is that it supports declarative changes. The system behavior can be modified by composing descriptions of object graphs (in XML), thus avoiding the long cycle of compile-pack-deploy that is required when the changes are done in code (in Java).

The MODELTALK framework is complemented by tool support [11]. An Eclipse [8] plug-in for MODELTALK provides developers with the same advanced level of environment look-and-feel for modeling as they are accustomed to with programming.

***Outline***    The rest of the paper is structured as follows. Section 2 briefly reviews dependency injection by example, comparing a code driven to a model driven approach. In Section 3 we describe the high level architecture of MODELTALK and show how modeling and coding are integrated to form a model driven development framework. In Section 4 we illustrate meta-modeling with MODELTALK. Assessment of the MODELTALK framework is brought in Section 5.

## 2.  Model Driven Dependency Injection

In this section, we illustrate the concept of code driven dependency injection in the Spring [21] framework. We then contrast dependency injection in Spring with the concept of model driven dependency injection in MODELTALK.

### 2.1  Code Driven Dependency Injection in Spring

In Spring, the developer starts the development iteration cycle by working on the Java implementation. Instances (*beans*, in Spring's terminology) are then defined to customize the implementation. As an example, consider the UML domain model for an HTTP client system depicted in Figure 1.[1] The Java class `HTTP_Client` (Listing 1) provides a `sendReceive` method for sending HTTP requests. The class has three private instance variables: `numberOfRetries` and `timeout` are used for configuring its communication handling policy; `URL` is used for configuring the Internet address of the resource to be accessed.

An XML bean in Spring is a description of an object graph. It is instantiated into Java objects at runtime. The XML excerpt in Listing 2 shows how one might use beans in Spring to customize the `HTTP_Client` class:

1. `RobustHTTP_Client` defines an abstract instance [21] of `HTTP_Client` with high numerical values for `timeout` and for `numberOfRetries`.

---

* This research was supported in part by the *Israel Science Foundation (ISF)* under grant No. 926/08 and by the office of the chief scientist of the Israel Ministry of Industry Trade and Labor.

[1] The UML diagrams are for illustration only. Models in MODELTALK are expressed in XML.

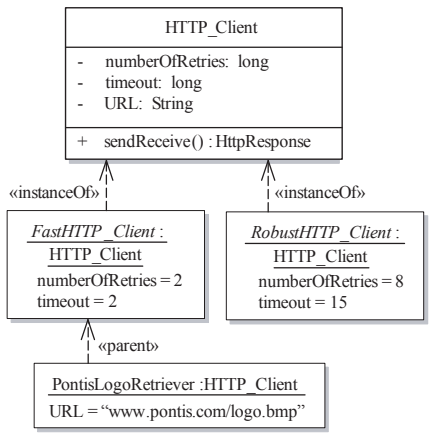

**Figure 1.** Domain model

```
public class HTTP_Client {
    private long numberOfRetries = 0;
    private long timeout = 0;
    private String URL = null;

    public void setNumberOfRetries(long number) {
        this.numberOfRetries = number;
    }
    public void setTimeout(long timeout) {
        this.timeout = timeout;
    }
    public void setURL(String URL) {
        this.URL = URL;
    }
    public HttpResponse sendRecieve() {
        HttpResponse result = null;
        //business logic
        return result;
    }
}
```

**Listing 1.** Class implementation in Java

```
<bean id="RobustHTTP_Client" class="HTTP_Client"
        abstract="true">
    <property name="numberOfRetries" value="8"/>
    <property name="timeout" value="15"/>
</bean>
<bean id="FastHTTP_Client" class="HTTP_Client"
        abstract="true">
    <property name="numberOfRetries" value="2"/>
    <property name="timeout" value="2"/>
</bean>
<bean id="PontisLogoRetriever" class="HTTP_Client"
        parent="FastHTTP_Client">
    <property name="URL"
            value="www.pontis.com/logo.bmp"/>
</bean>
```

**Listing 2.** XML beans in Spring

```
public static void main(String[] args)
        throws MalformedURLException {
    GenericApplicationContext context=getSpring();
    HTTP_Client httpClient =
        (HTTP_Client)context.getBean("PontisLogoRetriever");
    HttpResponse response = httpClient.sendReceive();
    //...
}
```

**Listing 3.** Client code in Java

2. `FastHTTP_Client` defines an abstract instance of `HTTP_Client` with low numerical values for `timeout` and for `numberOf-Retries`.

3. `PontisLogoRetriever` defines a *concrete* instance of `HTTP_-Client` by specializing `FastHTTP_Client` with the location for the logo bitmap.

This form of customization works for simple as well as for arbitrary complex object graphs. For simplicity, the example illustrates the use of Spring for configuring properties of primitive types. Generally, however, the injected values may also be instances of user defined classes.

Lastly, the Java excerpt in Listing 3 shows how a client code uses the Java factory to instantiate an `HTTP_Client` with the desired configuration.

### 2.2 Model Driven Dependency Injection in MODELTALK

In MODELTALK, the developer starts the development iteration cycle by working on the *model*. MODELTALK uses the notion of a class definition in the model. Model class definitions are the primary source in which the constraints for the XML beans and for the structure of the implementation code are defined. These constraints are reflected in the development tool immediately, providing the developer with full support for auto-completion, consistency checking, and so on.

The XML excerpt in Listing 4 is the model class definition of `HTTP_Client` (and its three model instances) in MODELTALK.[2] The definition informs the modeling tool about the existence of this class. This is in contrast to Spring, where one must have the Java class itself available. MODELTALK uses property name tags to provide domain specific syntax.

Model driven dependency injection enhances the safety of the declarative change process. Class definitions in the model constrain the model objects, leading to early detection of errors. Changes to the model can be applied to the runtime system with a higher degree of confidence than in Spring, since they undergo consistency checking.

In the next section we explain our model driven approach in more detail.

## 3. The MODELTALK Concept

The high-level architecture of MODELTALK (Figure 2) is similar to the general architecture of integrated development environments (IDEs). The source code processors (Figure 2B) are replicated to provide similar processors for modeling (Figure 2A). The architecture is implemented in an extensible IDE (Eclipse [8]) to yield an integrative model driven development environment.

### 3.1 Model Sources

In MODELTALK, model source files are textual and they are managed by the IDE just as other source files. The model sources com-

---

[2] We use a simplified dialect of MODELTALK concrete syntax.

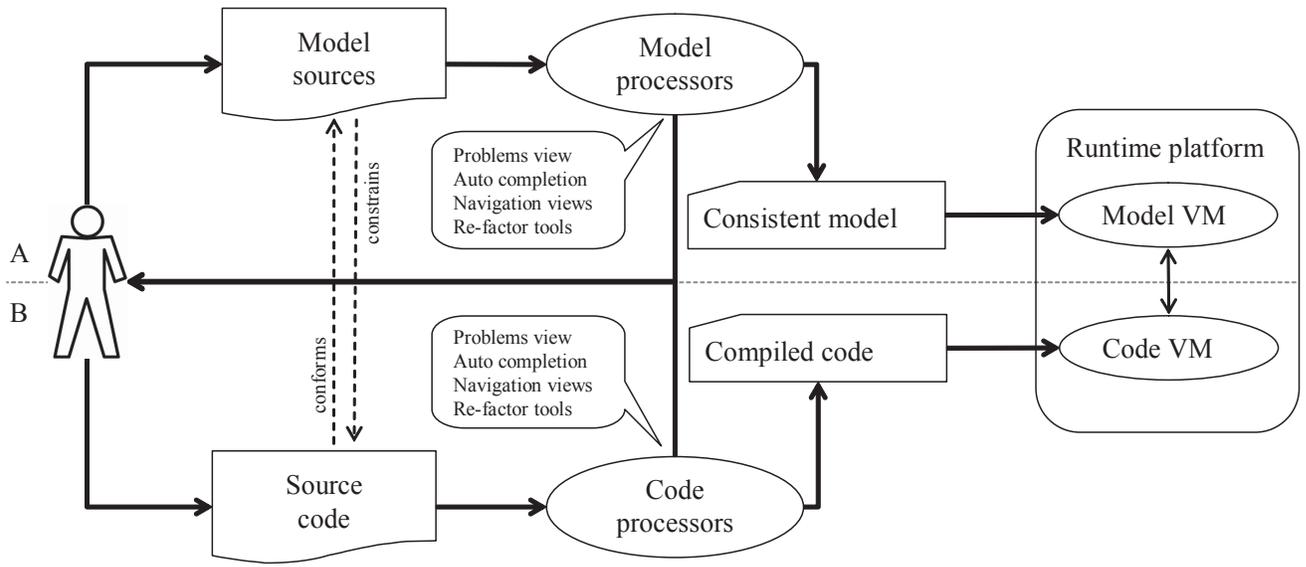

**Figure 2.** High level architecture of MODELTALK: (A) model processors; (B) code processors

```
<bean id="HTTP_Client" class="Class">
  <properties>
    <property>
      <name>numberOfRetries</name>
      <type>Long</type>
      <description>Number of retries</description>
    </property>
    <property>
      <name>timeout</name>
      <type>Long</type>
      <description>Timeout in seconds</description>
    </property>
    <property>
      <name>URL</name>
      <type>String</type>
      <description>The target URL</description>
    </property>
  </properties>
</bean>
<bean id="RobustHTTP_Client" class="HTTP_Client"
      abstract="true">
  <numberOfRetries>8</timeout>
  <timeout>15</timeout>
</bean>
<bean id="FastHTTP_Client" class="HTTP_Client"
      abstract="true">
  <numberOfRetries>2</timeout>
  <timeout>2</timeout>
</bean>
<bean id="PontisLogoRetriever" class="HTTP_Client"
      parent="FastHTTP_Client">
  <URL>www.pontis.com/logo.bmp</URL>
</bean>
```

**Listing 4.** Domain model in MODELTALK

prise instances, classes and metaclasses. A domain specific modeling language (DSML) [20] is formed by defining metaclasses and classes. In a typical scenario, the domain expert in the development team defines a DSML. Domain users then define models in this DSML. Since the modeling tools rely on class defined in the model rather than in the code, the modeling activity does not depend on the existence of implementation code.

### 3.2 Model Compiler

The model compiler is one of the model processors in the MODELTALK framework (Figure 2). It is implemented as an Eclipse builder plug-in. The compiler implements a dependency analysis algorithm to support incremental compilation.

Upon a change to the model, the compiler is invoked to perform cross-model validation. Object graphs in the model are validated against the corresponding model class definitions. This activity is analogous to how a compiler reports syntactical and certain semantical errors. Since MODELTALK is a meta-level system, classes, too, undergo similar validation checks. Cross-checks are necessary because a change in one model element might invalidate other model elements (possibly in other model files).

The model compiler also checks the conformance of the Java sources to the model [18]. When developing the code classes, the tool verifies conformance of the code structure to the model class definitions. Mismatches are reported as errors in the IDE standard problems view.

### 3.3 Model VM

The model VM is the runtime component of MODELTALK, which is analogous to the JVM. Its primary responsibility is to manage the relationships between model elements and Java elements. This includes object graph instantiation and a reflection API [14].

The model VM implements a dependency injection mechanism. When a client requests a model instance, the Model VM finds the corresponding Java class, instantiates it, and injects the model property values into the Java instance variables. This is applied recursively for injected value of a complex type. The Model VM algo-rithm for mapping model classes to Java classes permits "holes," i.e., a model class without a Java counterpart. In such a case, the class is called *declarative* and mapped instead to the superclass in the Java model. This adaptability enables to make changes to the model at runtime without needing to also change the Java model.

When a client requests a model class, the Model VM follows the same routine, thus enabling runtime modifications to the model. This is possible because MODELTALK's meta-meta-model itself is implemented in MODELTALK.

### 3.4 The User Experience in Modeling

The modeling user experience is similar to the user experience in programming. We use a commercial third-party XML editor as our model editor. The model editor provides auto-completion based on XML schema (XSD) generated from the model by the model compiler. Model elements are maintained in multiple source files arranged in folders by XML namespaces.

Model compilation is incremental, providing the user with short response time. Model compilation is done across all model files. Errors are reported to the developer using the standard IDE problems view. The developer can navigate to the erroneous model element by double clicking on the error [1].

The Eclipse plug-in provides numerous views of the model (e.g., type hierarchy) and provides navigation capabilities both between model elements themselves and between the model elements and corresponding Java elements. In addition, the plug-in provides refactoring facilities (e.g., rename) that propagate the changes to the Java source as well. Model source files are managed in a central repository (CVS) as other source files.

## 4. Meta-modeling with MODELTALK

In this section we illustrate the domain specific modeling capabilities of MODELTALK by extending the HTTP client example presented in Section 2. Suppose we would like to cache data in order to reduce network traffic and to improve the overall response time. Lets assume the application uses HTTP_Client to retrieve different kinds of data: pictures, news, stock quotes, etc. Obviously, various kinds of data require different caching policies. For example, pictures can be cached for longer periods than news, while stock quotes shouldn't be cached at all.

### 4.1 Declarative Classes

Implementing the caching code in Java in each of these classes would require the expertise of a Java developer. Instead, we can define a metaclass MetaCache with a cache property of type CacheManager (Figure 3 and Listing 5). The CacheManager provides cache management services at runtime. The methods getFromCache and putInCache are defined in the CacheManager class in Java. For brevity, the CacheManager and StandardCache classes are not shown in the listing.

We now make the HTTP_Client class an instance of MetaCache. The sendReceive method in HTTP_Client may then use the MetaCache metaclass to access the cache (Listing 6). We can further define specific HTTP client classes, PictureRetriever, NewsRetriever, and StockQuoteRetriever (Figure 4 and Listing 7), by subclassing HTTP_Client. Note that Picture-Retriever, NewsRetriever, and StockQuoteRetriever are *declarative* (without a counterpart in Java).

### 4.2 Customizing a Metaclass

Next, we enhance the example to demonstrate how architectural definitions are enforced by MODELTALK. Suppose our application needs to display bank account balances that are also retrieved using HTTP. Since bank account information is private, its confidentiality should be kept. We therefore have to

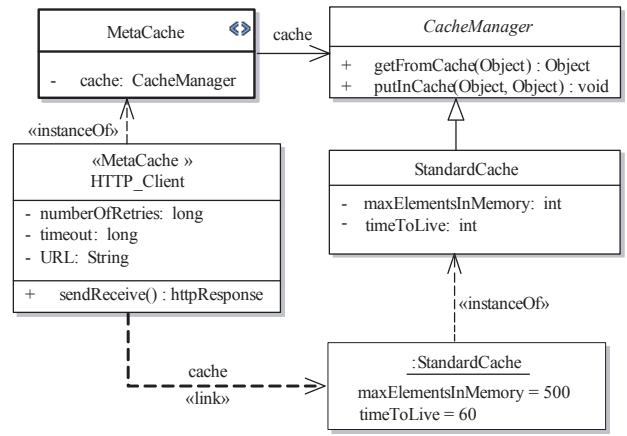

**Figure 3.** Domain model with a custom metaclass

```
<bean id="MetaCache" class="Class" parent="Class">
  <properties>
    <property>
      <name>cache</name>
      <type>CacheManager</type>
      <description>Caches the result</description>
    </property>
  </properties>
</bean>
<bean id="HTTP_Client" class="MetaCache">
  <cache class="StandardCache">
    <timeToLive>60</timeToLive>
    <maxElementsInMemory>500</maxElementsInMemory>
  </cache>
  <properties>
  ...
  </properties>
</bean>
```

**Listing 5.** MODELTALK sources with a custom metaclass

```
public HttpResponse sendRecieve() {
  MetaCache myMetaclass =
      (MetaCache)Kernel.instance().getClass(this);
  HttpResponse result =
      myMetaclass.getCache().getFromCache(getURL());
  if (result == null) {
    // do the business logic using timeout & numberOfRetries
    myMetaclass.getCache().putInCache(getURL(), result);
  }
  return result;
}
```

**Listing 6.** Accessing a metaclass in Java

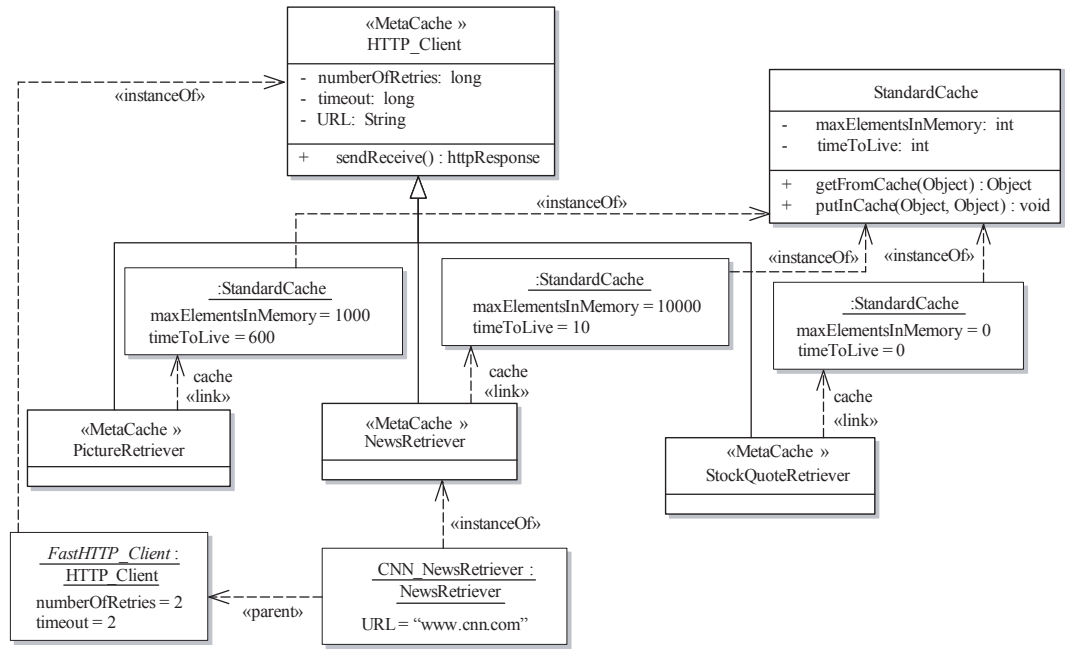

**Figure 4.** Expanding the domain model with different types of resources

```
<bean id="PictureRetriever" class="MetaCache" parent="HTTP_Client" declarative="true">
  <cache class="StandardCache">
    <timeToLive>600</timeToLive>
    <maxElementsInMemory>1000</maxElementsInMemory>
  </cache>
</bean>
<bean id="NewsRetriever" class="MetaCache" parent="HTTP_Client" declarative="true">
  <cache class="StandardCache">
    <timeToLive>10</timeToLive>
    <maxElementsInMemory>10000</maxElementsInMemory>
  </cache>
</bean>
<bean id="StockQuoteRetriever" class="MetaCache" parent="HTTP_Client" declarative="true">
  <cache class="StandardCache">
    <timeToLive>0</timeToLive>
    <maxElementsInMemory>0</maxElementsInMemory>
  </cache>
</bean>
<bean id="CNN_NewsRetriever" class="NewsRetriever" parent="FastHTTP_Client">
  <URL>www.cnn.com</URL>
</bean>
```

**Listing 7.** The resource retrieving model in MODELTALK

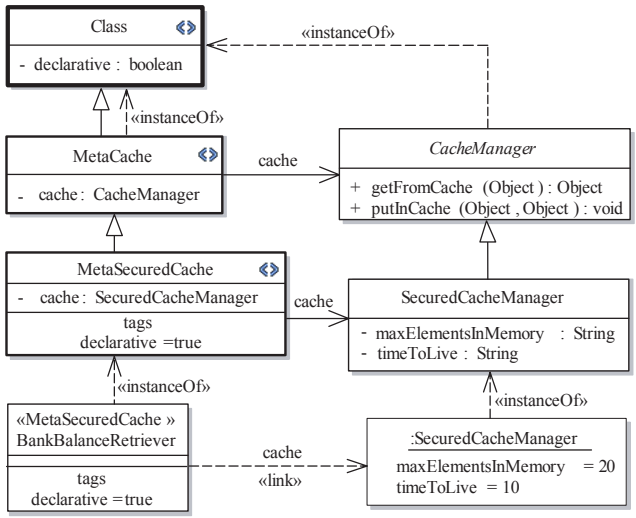

**Figure 5.** Expanding the domain model with secured cache

```
<bean id="MetaSecuredCache" class="Class"
      parent="MetaCache" declarative="true">
  <properties>
    <property>
      <name>cache</name>
      <type>SecuredCacheManager</type>
      <description>Provides secured caching
        capabilities</description>
    </property>
  </properties>
</bean>
<bean id="BankBalanceRetriever"
      class="MetaSecuredCache"
      parent="HTTP_Client" declarative="true">
  <cache class="SecuredCacheManager">
    <timeToLive>10</timeToLive>
    <maxElementsInMemory>20</maxElementsInMemory>
  </cache>
</bean>
```

**Listing 8.** The secured cache model in MODELTALK

make sure that only a secured cache manager is used in such cases. To achieve this we define `MetaSecuredCache` that extends `MetaCache` and overrides the type of the *cache* property to `SecuredCacheManager`, which is a subclass of `CacheManager` (Figure 5 and Listing 8). `BankBalanceRetriever` is defined as instance of `MetaSecuredCache` and is therefore required to supply an instance of a `SecuredCacheManager` for its cache property; otherwise, a model compilation error will be issued. For the sake of brevity, we do not show the corresponding changes in the Java code.

Custom metaclasses [5, 6] define class level properties, which constrain the domain classes using domain terminology. Unlike class level members in Java (static members), metaclass properties allow subclasses to have different values for the metaclass properties.

The model contains objects, classes and metaclasses in a single type system. The same injection mechanism that works on objects is applied to classes as objects of their metaclasses. Since Java does not support metaclass extensibility, the metaclasses in the model are mapped to regular Java classes and the model VM manages the *instance-of* relation in the runtime system.

## 5. Assessment

The MODELTALK development platform has been used at Pontis by a team of over 20 developers for the last two years. Numerous customer projects were developed, delivered and deployed successfully, satisfying requirements for hundreds of transactions per second. In this section we describe our subjective observations from using MODELTALK in a commercial product development environment.

Since the inception of the platform, we have been continuously running cage measurements in order to indicate the platform's level of adoption within the R&D organization. Currently, the model contains approximately 4800 classes, of which 200 are metaclasses. There are tens of thousands of instances and 275K lines-of-code in XML. The average Depth of Inheritance (DIT) of model classes is 4.75. The XML schema of the application model (generated XSD) is 200K lines-of-code. 90% of the application Java source code is governed by the model (i.e., the classes are defined in the model). 82% of the source lines-of-code in customization projects are declarative.

### 5.1 Developer Perspective

Developers give very positive feedback, mostly concerning the dramatic improvement in cycle times. An incremental model change takes no more than a few seconds on large models, compared to minutes, at best, in generative MDD (e.g., [23, p. 256] and [23, p. 261]).

Users appreciate the Java-like usability of MODELTALK and the fact that modeling and programming are done in a single, integrated environment. Specifically, the developers mention the ability to work on "broken models." For example, when changing a property name in a class, instances of the class become invalid. The MODELTALK environment allows the developer to continue modeling although part of the model is temporarily in an inconsistent state.

Users get accustomed to formal modeling very quickly and rely on the model compiler to enforce architectural constraints. Users complained about the tedious, manual work in writing the structural part of the Java code, especially getters/setters. To address this, we extended MODELTALK with some code generation capabilities, which is outside the scope of this paper. The generation of getters and setters provides Java type-safety, but is strictly optional, because the model is fully reflective.

Users also complain about the lack of diagramming capabilities and interoperability with UML modeling tools. This is a topic we plan to address in future work.

### 5.2 Organization Perspective

An organization considering adopting a similar approach should take into account a substantial initial investment for building the infrastructure and tools. In our case, the investment was more than 10 man years. In addition, the ongoing maintenance of the development environment must be considered. Another barrier is the inherent complexity of such an approach. Most developers are not familiar with meta-modeling and need extensive training. Moreover, application implementation tends to be highly abstract and generic, which requires highly talented individuals.

However, once such a development environment is in place, there are tangible benefits for the organization. First and foremost,

all the advantages of MDD and DSML apply [23, 22]. Especially, time-to-market and the cost of producing customized products drop significantly. Second, the ability to deploy a compiled model directly to a running system (when changes in Java are not required) creates a much shorter delivery route.

# 6. Conclusion and Future Work

Software solutions in the telecommunications industry typically require massive customization. In order to reduce the cost and time-to-market of creating customized products, we developed MODELTALK, a domain specific model driven framework, and used the framework in product development.

An early implementation of MODELTALK was based on a generative architecture centric MDSD approach [23]. The advantages of the model centric approach were evident. However, when the model evolved to thousands of classes, developers started to complain about the long development cycles (several minutes for each incremental change). This stemmed from the large amounts of generated Java code that had to undergo compilation, packaging and deployment to the J2EE application server.

In this paper we presented the new version of MODELTALK, which is based on an executable model, dependency injection, and meta-modeling; and complemented by a model compiler and tooling. Together these provide an enhanced user experience for the modeling process, similar to the programming user experience in modern IDEs.

The meta-level capabilities of MODELTALK are used by developers to create custom types of classes, fields and methods resulting in a domain specific modeling language. There is also support for resolving crosscutting concerns at the model level [3, 2, 4, 7, 13].

We are currently working on enhancing MODELTALK with runtime adaptability, i.e., the ability of non-programmers to change the model of a production system [9, 16]. The interpretive nature of the MODELTALK platform provides a sound basis for achieving this by combining an interpretive approach with metaclass extensibility [5].

## Acknowledgments

We thank Zvi Ravia and Shai Koenig for their valuable comments. We thank our colleagues in Pontis and partners in developing the MODELTALK platform: Shachar Segev, Moshe Moses and Assaf Pinhasi. Pontis Ltd. is a developer of online marketing automation solutions for the telecommunications market.